\documentclass{article}

\usepackage{PRIMEarxiv}

\usepackage[utf8]{inputenc} 
\usepackage[T1]{fontenc}    
\usepackage{hyperref}       
\usepackage{url}            
\usepackage{booktabs}       
\usepackage{amsfonts}       
\usepackage{nicefrac}       
\usepackage{lipsum}
\usepackage{fancyhdr}       
\usepackage{graphicx}       
\graphicspath{{media/}}     

\usepackage{amsmath}
\usepackage[nopatch]{microtype}

\usepackage{graphicx,tipa}
\usepackage{graphicx}                   
\usepackage[T1]{fontenc}
\usepackage{amsmath}                    
\usepackage{amssymb}                    
\usepackage{url}                        
\usepackage{hyperref}                   
\usepackage{multirow}                   
\usepackage{booktabs}                   

\usepackage{float}                      
\usepackage{cleveref}                   
\usepackage{color}                      
\usepackage{lipsum}                     
\usepackage{subcaption}                 
\usepackage{siunitx}                    
\usepackage{pdflscape}                  
\usepackage{ragged2e}
\usepackage{lineno}  
\usepackage{array}
\usepackage{caption}
\usepackage{tabularray}
\usepackage{multicol}
\usepackage{longtable}
\usepackage{colortbl}

\newcommand\arcsec{\mbox{$^{\prime\prime}$}}%
\newcommand\arcmin{\mbox{$^\prime$}}%

\title{A sky survey of ultraviolet sources observed through AstroSat's UVIT: A point source catalog

}

\author{
  Swagat Bordoloi, Rupjyoti Gogoi \\
  Department of Physics, Tezpur University, Napaam,  784028, Assam, India \\
   \And
  P. Shalima \\
  Manipal Centre for Natural Sciences, Manipal Academy of Higher Education, Manipal, 576104, Karnataka, India \\
   \And
  Jayant Murthy \\
 Indian Institute of Astrophysics, Bengaluru, 560034, Karnataka, India \\
}

\begin{document}
\maketitle

\begin{abstract}
 The Ultra Violet Imaging Telescope (UVIT) onboard India's first dedicated multiwavelength satellite \textit{AstroSat} observed a significant fraction of the sky in the ultraviolet with a spatial resolution of 1.4\arcsec. We present a catalog of the point sources observed by UVIT in the far ultraviolet (FUV; 1300-1800 \AA) and near ultraviolet (NUV; 2000-3000 \AA). We carried out astrometry and photometry of 428 field pointings in the FUV and 54 field pointings in the NUV band, observed in 5 filter bands in each channel respectively, covering an area of about 63 square degrees. The final catalog contains about 102,773 sources. The limiting magnitude(AB) of the F148W band filter, that has the largest number of detections is $\sim21.3$. For the NUV channel, we find the limiting magnitude at around $\sim23$. We describe the final catalog and present the results of the statistical analysis.
\end{abstract}


\section{Introduction}
The Indian Space Research Organisation (ISRO) launched \textit{Astrosat} \cite{Agrawal06} on September 28, 2015. It was  India's first dedicated multi-wavelength astronomy satellite and included 5 payloads: the Soft X-Ray imaging Telescope (SXT); the Large Area X-Ray Proportional Counters (LAXPC); the Cadmium Zinc- Telluride Imager (CZTI); the Scanning Sky Monitor (SSM); and the Ultra Violet Imaging Telescope (UVIT), the only ultraviolet (UV) telescope onboard. 

    There are two telescopes on the UVIT payload which simultaneously image the sky in the far ultraviolet (FUV: 1300-1800 \AA) in one telescope, and the near ultraviolet (NUV: 2000 - 3000 {\AA}) and visible (VIS: 3200-5300 \AA) in the second telescope, with the bands separated using a dichroic mirror. The field of view (FOV) is $\sim28\arcmin$ diameter with a spatial resolution of $\sim1.4\arcsec$. Specific bands are selected through a filter wheel on each telescope. The VIS data were intended to be used only for astrometric correction and have no scientific utility. The instrument has been described in Kumar et al\cite{kumar2012}.

An important product of a wide-field imaging mission is a point source catalog that may be correlated with catalogs in other wavelengths to yield a multi-wavelength picture of the sky. While numerous catalogs exist in the optical (e.g., \textit{Sloan Digital Sky Survey} (SDSS) and \textit{Panoramic Survey Telescope and Rapid Response System} (Pan-STARRS)) and infrared (e.g., \textit{2-Mass} Catalog), fewer options are available in the UV due to the need for space-based observations. An early attempt was the TD-1 catalog \cite{td1mission} from the TD-1 mission, a satellite operated by the \textit{European Space Research Organisation} (ESRO). The Ultraviolet Sky Survey Telescope onboard TD1 measured the absolute ultraviolet flux distribution of point sources between 1350\AA - 2550\AA, containing about 31,215 stars with a visual magnitude limit of $\sim$10 for unreddened early B type stars in its six-month observation period. The deepest UV catalog to date is from the \textit{Galaxy Evolution Explorer} (GALEX) \cite{galexmission2005,bianchi2017}, which was observed in two main bands (1350\AA-1750\AA and 1750\AA-2750\AA). The GALEX catalog includes data from three surveys: All-sky Imaging Survey (AIS), Medium Imaging Survey (MIS), and Deep Imaging Survey (DIS), covering approximately 25,000 sq. degrees of sky in its final data release. The AIS provides a magnitude limit of $\sim$20, while the MIS and DIS can detect sources up to magnitude limits of $\sim$23 and $\sim$25, respectively.

There exist catalogs of small portions of the sky observed by UVIT. \cite{leahy2020}Leahy et al. observed about 18 regions of the M31, and generated a catalog of $\sim$75,000 sources with a limiting magnitude of $\sim$23 in FUV CaF2-1 filter. \cite{devaraj2023}Devaraj et al. created a point source catalog from the observation of 3 overlapping fields in the outskirts of the SMC by the UVIT. The total number of sources detected was $\sim$11,241. The catalog provided information about their AB magnitude in 7 UVIT filters, of which 3 are in the FUV and 4 in the NUV. In our work, we cover a larger area of the sky with a higher number of field images, including a few fields of SMC.

UVIT has completed about $\sim${1700} observations (including multiple observations of a single field) after almost 9 years of observations and we believe that it is time to construct a point source catalog (\textit{'UVIT-cat'}). We begin with a sample of  {\bf 428 fields} in the FUV and {\bf 54 fields} in the NUV covering about $\sim63$ square degrees of the sky and plan to expand the catalog to the entire UVIT data set. In Section 2, we explain the details of the UVIT instrumentation and data collection; in Section 3, we discuss the analysis and results; and in Section 4, we give a summary of our catalog and future work.

\section{UVIT Instrument and Data}

    The Ultra Violet Imaging Telescope (UVIT) observed the sky in two UV and one visible band using two 38 cm telescopes. One telescope observed the far ultraviolet (FUV: 1300-1800 \AA) while the other split the light into the near ultraviolet (NUV: 2000 - 3000 {\AA}) and visible (VIS: 3200-5300 \AA) bands. There are three identical intensified microchannel plates with an aperture of 40mm, and only the photocathode differs as given in Kumar et al \cite{kumar2012}. A filter wheel on each telescope divided the spectral region into several bands (Table \ref{table:1}).  The UVIT image pixel size is {$\simeq$0.4168\arcsec} by {0.4168\arcsec.} The two UV detectors were operated in photon counting mode while the VIS channel was used only for attitude correction and was operated in integration mode.

    UVIT data are processed and archived at the Indian Space Science Data Centre \cite{astrosatmission} and may be downloaded from the \textit{Astrobrowse Archive}\footnote{\url{https://astrobrowse.issdc.gov.in/astro_archive/archive/Home.jsp}} in the following formats:
\begin{enumerate}
    \item \textbf{Level 0:} The Level 0 is the raw binary data as observed by the UVIT payload onboard the \textit{AstroSat}, along with its auxiliary data. This is sent to the AstroSat Data Centre at \textit{Indian Space Science Data Centre} (ISSDC) for further processing into Level 1 data.

    \item \textbf{Level 1:} The Level 1 data from the ISSDC is sent to the \textit{Payload Operation Centre} (POC) at the Indian Institute of Astrophysics (IIA), Bengaluru. The files in the Level 1 data are FITS binary table files organized according to the orbit number under a single top-level directory as described in Rahna et al.\cite{jude2}. 

    \item \textbf{Level 2:} These are the final products after processing the Level 1 data through the UVIT pipeline software. The final product is a FITS image file, containing the coordinate information and can be read by any common astronomical data processing program.
\end{enumerate}

\begin{table}
    \centering
    \resizebox{\linewidth}{!}{\begin{tabular}{cccccc}
         \hline
Filter & Filtername & Mean $\lambda$  & Zeropoint (ZP)    & passband \\
       &            & (\AA)         & (ZP)            & (nm)\\
 \hline
F148W  & CaF2-1     & 1481       & 18.097          & 125–179                     \\ 

F154W  & BaF2       & 1541       & 17.771          & 133–183                     \\ 

F169M  & Sapphire   & 1608       & 17.410          & 145–181                    \\ 

F172M  & Silica     & 1717       & 16.274          & 160–179                     \\

\hline 
N242W  & Silica-1   & 2418       & 19.763          & 194–304                    \\ 

N219M  & NUVB15    & 2196       & 16.654          & 190–240                 \\ 

N245M  & NUVB13   & 2447       & 18.452           & 220–265                   \\ 

N263M  & NUVB4     & 2632       & 18.146         & 220–265                    \\

N279N  & NUVN2     & 2792       & 16.416         & 273–288                  \\

\hline

    \end{tabular}}
    
    \caption{Filter details used for observations}
    \label{table:1}
\end{table}

\begin{figure*}
    \centering
    \includegraphics[width=\linewidth ]{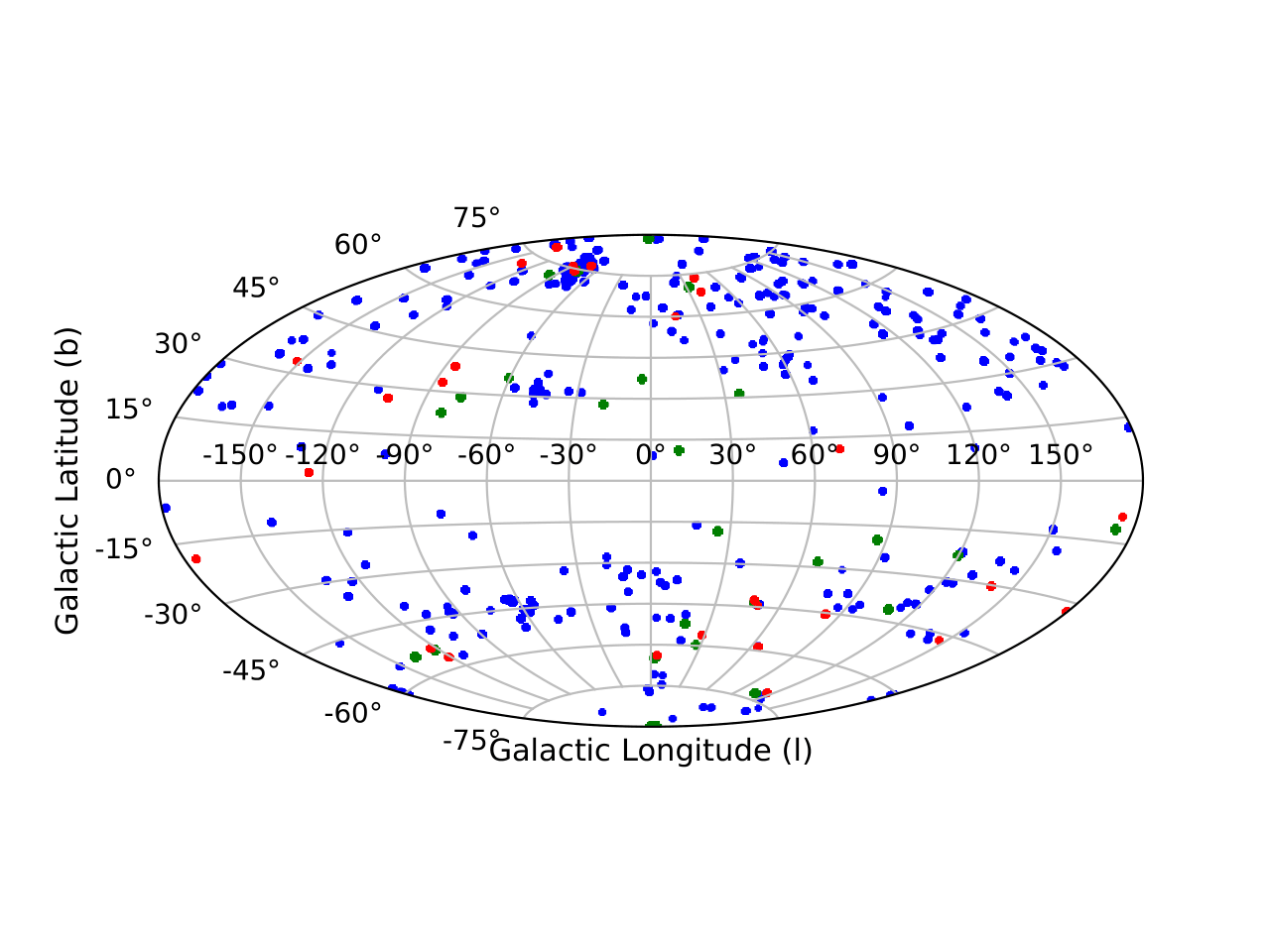}
    \caption{Area of sky covered by UVIT-cat. Blue: FUV fields only, Green: NUV fields only, and Red: Field images containing both FUV and NUV}
    \label{fig:sky_plot}
\end{figure*}

    As of 1st June 2024, there were a total of 1713 observations in the UVIT archive in both channels. The NUV channel ceased operation in March 2018, and only FUV data were taken since then. We have only used observations with an exposure time greater than 200 seconds. These are distributed in the sky as shown in Fig \ref{fig:sky_plot} and exclude the Galactic Plane, where there were no observations due to concerns about the sky brightness. The blue dots represent the images observed only through the FUV filter, while the green dots represent sources observed only through the NUV filter. We represent the pointings as red dots for field images observed through both FUV and NUV. 

\subsection{Astrometric Correction and Source Extraction}

    The astrometry in the Level 2 archival data file is off by several arcminutes, and each field has to be corrected individually. We first ran each image through \textit{Astrometry.net} by Lang et al.\cite{Dustinastrometry}, which used a set of stars (or galaxies) from a reference catalog (typically optical catalogs such as Tycho-2 and 2MASS) to solve for the astrometry in that image. This worked well for the NUV, where the sky looks similar to the optical, but the FUV sky is quite different from the visible sky, and we had to create a custom catalog based on GALEX FUV observations. There were still a few problematic fields, generally because there were too few stars in the field, which we corrected by matching stars by eye. The FITS files of the image headers having the updated World Coordinate system are in the GitHub repository \footnote{\url{https://github.com/swagatastro98/UVIT-cat.git}}. 

    We used SExtractor v2.25.0 \cite{bertinsextractor} to extract point sources from the astrometrically-corrected FITS files. We used the default parameters for extraction except as listed below:

   \begin{enumerate}   

    \item We changed DETECT\_MINAREA (the minimum number of adjacent pixels that have to be above a certain value for a detection) to 5 from 3 as per the PSF of the instrument.
        
    \item The value for DETECT\_THRESH was set to be 3 for the FUV and 5 for NUV images to minimize the number of false detections. Setting a smaller threshold value (e.g. 1.5 times background) increased the false detection counts of higher exposure images. Hence, we chose an optimal value for the detection threshold such that there is minimum detection of false positives across the entire dataset. 
    
     \item We used the pixel scale of UVIT (0.4168\arcsec) for PIXEL\_SCALE

    \item The zero point magnitude (MAG\_ZEROPOINT) of the instrument was varied for each filter, as per \cite{tandon2020additional}.
    \end{enumerate}
    
    The detailed parameters used for source extraction are listed in Table \ref{table:2}
    
    The source-extractor calculates the error in magnitudes by considering the error in the flux, which is given as:

    \begin{align}
    \sigma_{F} = \sqrt{\sigma^2_{photon}+ \sigma^2_{sky} + \sigma^2_{read}}  \label{eq1}
    \end{align}
    
    where, $\sigma_{photon}$ is the photon noise, $\sigma_{sky}$ is the background noise and $\sigma_{read}$ is the readout noise associated with the detector.
    
    The photon noise can be represented as the square root of flux in photon counts (F) from the source as:
    \begin{align}
        \sigma_{F} = \sqrt{F}
    \end{align}

    The magnitude error, $\Delta m$, is calculated using the flux error value from equation \eqref{eq1} as:

    \begin{align}
    \Delta m &= \frac{2.5}{\ln10}(\frac{\sigma_{F}}{F})
    \end{align}

The detection threshold parameters and the error calculation algorithm can be found in the manual at Source-extractor official website\footnote{\url{https://www.astromatic.net/software/sextractor/}}. 

\begin{table}

        \begin{tabular}{cc}
            \hline
            Parameters   & Values  \\ 
            
            \hline
            DETECT\_MINAREA &  5 \\
            DETECT\_THRESH (FUV/NUV) & 3/5 \\ 
            ANALYSIS\_THRESH &  3 \\
            FILTER &  default.conv \\
            DEBLEND\_NTHRESH &  32 \\
            DEBLEND\_MINCONT &  0.005 \\
            PIXEL\_SCALE & 0.41657 \\
            BACK\_TYPE & AUTO \\
            BACK\_SIZE & 8 - 256 \\
            BACK\_FILTERSIZE & 1 - 3 \\
            CLEAN & Y \\
            CLEAN\_PARAM & 0.1 \\
            \hline
        \end{tabular}
        \caption{Source-Extractor default parameters}
        \label{table:2}
    \end{table}
    
\subsection{Merging of duplicate sources}

We combined the point source list from each observation into a single catalog.
Any detection within 1.5\arcsec of another was merged into a single source. A single binary FITS binary table is created from all observations of a single field, regardless of detector or filter. If multiple detections of the same source were present in the same filter, we took the mean of the magnitudes and flux. If multiple detections of a source were present in different filters, all the information is condensed into a single row in the FITS table. The number of sources having entries in FUV and NUV are given in Table \ref{table:total_sources} and \ref{table:total_sources_nuv}.

\begin{table}
        
        \resizebox{\linewidth}{!}{
        \begin{tabular}{ccccc}
            \hline
            Filter-slot & Filter & Filter-name & Number of fields   & Number of sources  \\ 
            
            \hline
            F1 & F148W & CaF2-1   & 190 & 44,289 \\
            F2 & F154W & BaF2     & 157 & 16,161 \\
            F3 & F169M & Sapphire & 60  & 6,739 \\
            F5 & F172M & Silica   & 17  & 2,690 \\
            F7 & F148Wa & CaF2-2   & 4   & 1,574 \\
            \hline
            Total &    & & 428 & 71,453 \\
            \hline
        \end{tabular}}
        \caption{Total number of sources through FUV filter}
        \label{table:total_sources}
    \end{table}

\begin{table}
        
        \resizebox{\linewidth}{!}{
        \begin{tabular}{ccccc}
            \hline
            Filter-slot & Filter & Filter-name & Number of fields   & Number of sources  \\ 
            
            \hline
            N1 & N242W & Silica-1   & 26 & 24,562 \\
            N2 & N219M &NUVB15     & 7 & 546 \\
            N3 & N245M &NUVB13 & 15  & 9,965 \\
            N5 & N263M &NUVB4   & 3  & 1,942 \\
            N6 & N279N &NUVN2   & 3   & 3,238 \\
            \hline
            Total &   &  & 54 & 40,253 \\
            \hline
        \end{tabular}}
        \caption{Total number of sources through NUV filter}
        \label{table:total_sources_nuv}
    \end{table}

\section{Result and Analysis}

    \subsection{Image to Catalog Script}

    We have modified \textit{jude\_call\_astrometry} procedure of JUDE (Jayant's UVIT Data Explorer) \cite{jude2017, jude2} to process the Level 2 UVIT images, correct the astrometry, and output a FITS binary table catalog file for each image. The columns are listed in Table \ref{table:summary_table}. We then merged all the field catalogs into a single file (UVIT-cat). We have included all instances of different observations of the same field. 

    The UVIT catalog without any merging had about $\simeq117,598$ sources, of which $14,825$ were duplicates. We merged the duplicate rows to come up with the UVIT-cat catalog that contains \textbf{102,773} NUV and FUV sources, observed through different filters of UVIT. 

    The matching algorithm merges all sources within a 1.5\arcsec radius to give us a final merged catalog. The algorithm finds all sources within the radius by flagging them as duplicates, extracts the magnitudes, fluxes, and their errors from each row in a single filter, and computes the average, which is taken as the final photometry, and merged into a single row. If duplicate sources are found in separate filters, all the rows are merged into a single row with photometric entries in separate columns.

    Many field images with extended sources like galaxies, nebulas, and densely crowded fields, are not yet included, and we intend to put forward updated versions of the catalog by including more UVIT field images, both in FUV and NUV.

\begin{table}
	\centering

        \resizebox{\linewidth}{!}{
	\begin{tabular}{cc} 
 
		\hline
		Tags & Columns \\
		\hline
		
		RA & Right ascension of the centroid position of the source \\
		DEC & Declination of the centroid position of the source \\
            N1$\_$M$\_$A & NUV Kron-like elliptical aperture magnitude (F1 filter)\\
            N1$\_$MER$\_$A & NUV RMS error for elliptical aperture magnitude (F1 filter)\\
            N1$\_$M$\_$I & NUV Isophotal magnitude (F1 filter)\\
            N1$\_$MER$\_$I & NUV RMS error for isophotal magnitude (F1 filter)\\
            FLAGS & Extraction flags \\
            N1$\_$F$\_$A & NUV Flux within a kron-like elliptical aperture (F1 filter)\\
            N1$\_$FER$\_$A & NUV Flux error within a kron-like elliptical aperture (F1 filter)\\
            A$\_$IMAGE & Profile RMS along major axis \\
            B$\_$IMAGE & Profile RMS along minor axis \\
            A$\_$WORLD & Profile RMS along the major axis (world units) \\
            B$\_$WORLD & Profile RMS along the minor axis (world units) \\
            X$\_$IMAGE & Object position along X-axis (in pixels) \\
            Y$\_$IMAGE & Object position along Y-axis (in pixels) \\
            X$\_$WORLD & Object position along X-axis (in degrees) \\
            Y$\_$WORLD & Object position along Y-axis (in degrees) \\
            ERRA$\_$IMG & RMS position error along major axis \\
            ERRB$\_$IMG & RMS position error along minor axis \\
            ERRA$\_$WLD & World RMS position error along major axis \\
            ERRB$\_$WLD & World RMS position error along minor axis \\
            XY$\_$WORLD & Covariance between X-world and Y-world \\
            FWHM$\_$WLD & FWHM of the object assuming a Gaussian core (in degrees)\\
            FWHM$\_$IMG & FWHM of the object assuming a Gaussian core (in pixels)\\
            CLS$\_$STR & Star/Galaxy classifier \\
            ELLIP & Ellipticity; 1 - B$\_$image/A$\_$image \\
            THEJ2000 & Position angle (J2000) \\
            KRON$\_$RAD & Kron apertures in units of A or B \\
            GAL$\_$RA & Galactic right ascension \\
            GAL$\_$DEC & Galactic declination \\
            DIST$\_$FOV & Distance of the object from the center of FOV \\
            F1$\_$M$\_$A & FUV Kron-like elliptical aperture magnitude (F1 filter)\\
            F1$\_$MER$\_$A & FUV RMS error for elliptical aperture magnitude (F1 filter)\\
            F1$\_$M$\_$I & FUV Isophotal magnitude (F1 filter)\\
            F1$\_$MER$\_$I & FUV RMS error for isophotal magnitude (F1 filter)\\
            F1$\_$F$\_$A & FUV Flux within a kron-like elliptical aperture (F1 filter)\\
            F1$\_$FER$\_$A & FUV Flux error within a kron-like elliptical aperture (F1 filter)\\
            
		\hline
	\end{tabular}}
	\caption{UVIT-cat catalog summary}
	\label{table:summary_table}
\end{table}

  \subsection{Completeness and Analysis of the Catalog}

    The number of fields along with their respective number of sources observed through each FUV and NUV filter of UVIT is given in Table \ref{table:total_sources} and \ref{table:total_sources_nuv} respectively. We find the highest number of observations and detections in the first filter of both NUV and FUV channels. 44,289 sources were found in the F148W filter (CaF2-1; F1) of FUV channel out of the 190 observations. A similar (157) number of observations were made through the filter F154W (BaF2; F2), but significantly less number of detections were made (16,161) as shown in Table \ref{table:total_sources}. The number of sources detected is about $\simeq24,562$ as detected in the 26 fields of the NUV N242W filter. The next set of filters (N219M, N245M, N263M, and N279N) detected about 546, 9,965, 1,942, and 3,238 sources respectively. We found that the highest exposure time images were mostly observed through the F1 filter of FUV and NUV compared to the other filters, and this accounts for the highest number of detections of the first filter in each channel.

\begin{figure}
    \centering
    \includegraphics[width=\linewidth ]{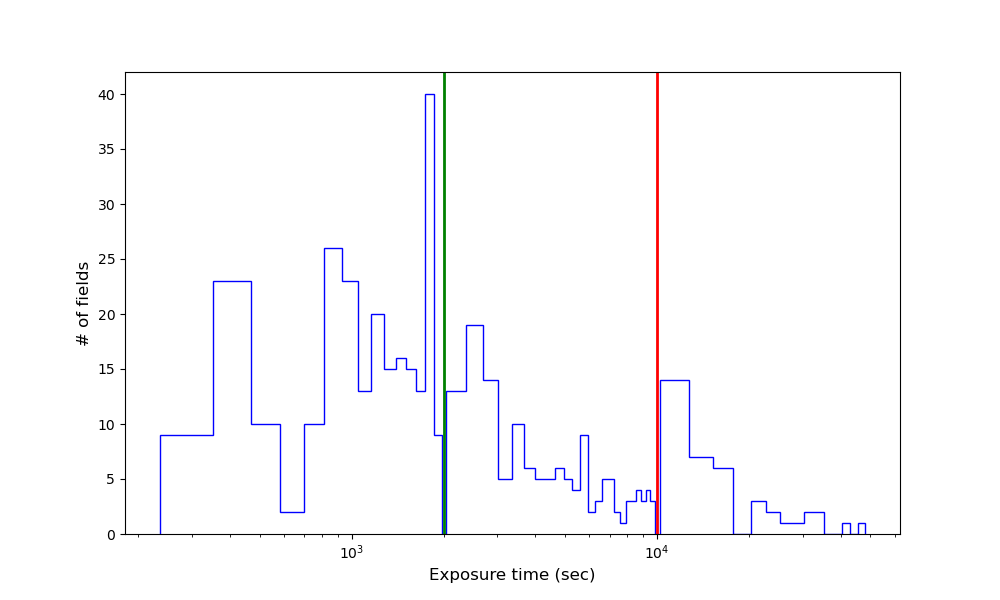}
    \caption{Exposure time histogram of 3 bins in FUV band, separated by (a)green: 2,000s and (b)red: 10,000s}
    \label{fig:Exposure-time-histogram}
\end{figure}

\begin{figure*}
        \centering
        \begin{multicols}{2}
            \subcaptionbox{Exposure time histogram of NUV images\label{fig:Exposure time NUV}}{\includegraphics[width=\linewidth]{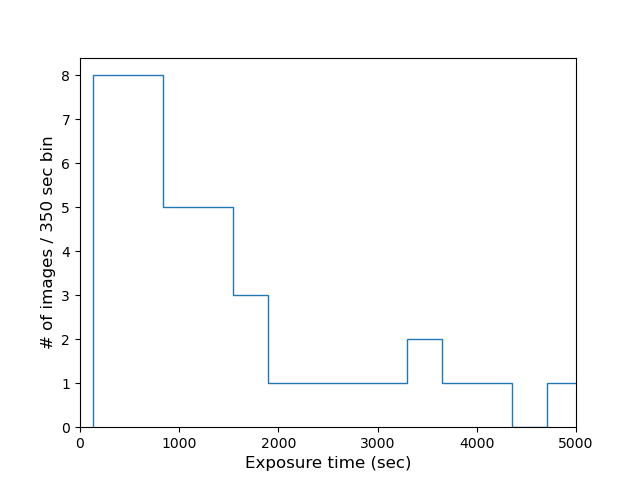}}\par 
            \subcaptionbox{AB Magnitude vs Source count distribution of NUV\label{fig:nuv source count}}{\includegraphics[width=\linewidth]{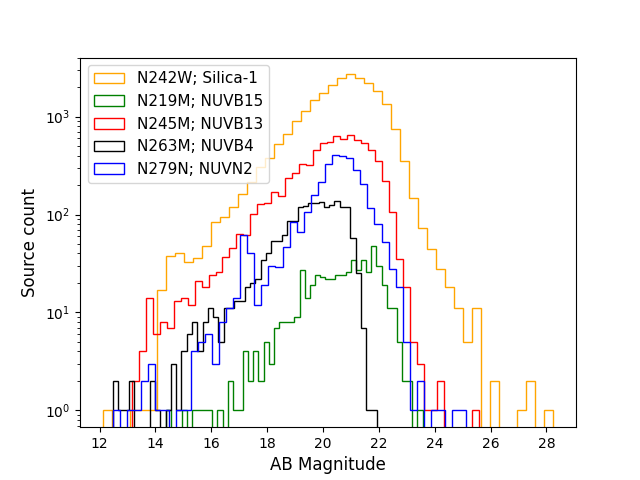}}\par
            
        \end{multicols}
    \caption{(a) Histogram of exposure times in NUV band (b) AB magnitude vs Source count distribution of NUV filters}
    \label{fig:Magnitude Distribution}
    \end{figure*}

    The images that were analyzed had a large variation of exposure time (from $\sim400$s to more than 45,000s), influencing the catalog's completeness. To facilitate statistical studies, we divided the images into three bins based on exposure times. The first bin contained images with 2,000s or less exposure timing. The second bin ranged from 2,000 to 10,000s, and any image with exposure time greater than 10,000s was put into the third bin. The bins contained 246, 142, and 40 FUV images respectively. As shown in Fig \ref{fig:Exposure-time-histogram}, the magnitude limit will correspond to the median value of each of the three bins. For the first bin, we find the typical exposure of FUV images to be $\sim1798$s. The second and third bins had the median exposure time at $\sim2,500$s and $\sim11,000$s respectively. The bins are separated with two vertical lines, green and red at the 2000s and 10,000s respectively. The histogram of the exposure time of NUV images is shown in Fig \ref{fig:Exposure time NUV}, and we find out the median value to be $\sim400$s.

\begin{figure*}
        \centering
        \begin{multicols}{3}
            \subcaptionbox{Exposure time <2,000s\label{fig:Source count bin1}}{\includegraphics[width=\linewidth]{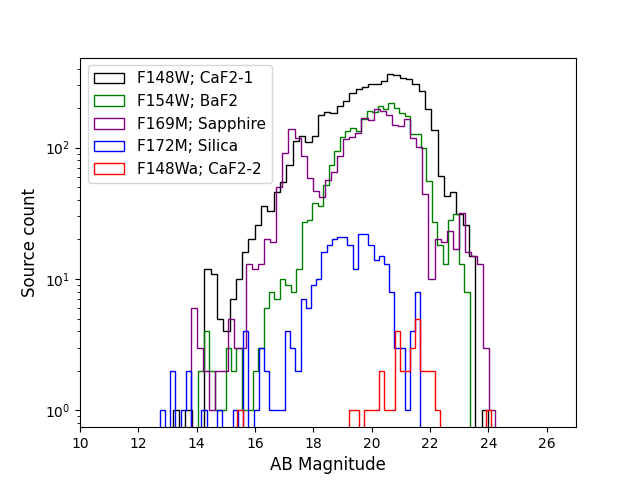}}\par 
            \subcaptionbox{Exposure time >2,000s and <10,000s\label{fig:Source count bin2}}{\includegraphics[width=\linewidth]{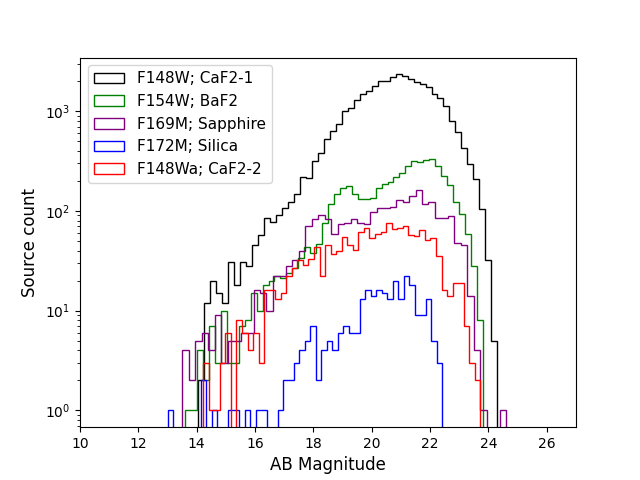}}\par
            \subcaptionbox{Exposure time >10,000s\label{fig:Source count bin3}}{\includegraphics[width=\linewidth]{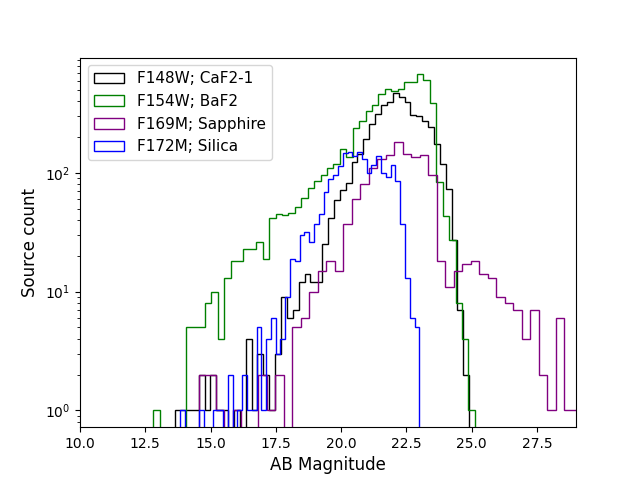}}\par
            
        \end{multicols}
    \caption{AB magnitude vs Source count distribution for 3 bins in FUV band}
    \label{fig:Magnitude distribution plot FUV}
    \end{figure*}

The distribution of the AB magnitude of the detected sources in each of the 3 bins of the FUV channel and NUV channel is shown in Fig \ref{fig:Magnitude distribution plot FUV} and Fig \ref{fig:nuv source count}. The plots are shown for images observed via different filters of the FUV and NUV channels of UVIT.

    We find the peak of the source count distribution as 20.527, 20.81, and 21.98 for the F1 filter in the three respective bins of exposure times as shown in Fig \ref{fig:Source count bin1}, \ref{fig:Source count bin2} and \ref{fig:Source count bin3}. The peak magnitude distribution for the remaining filters of UVIT in FUV is tabulated in Table \ref{table:peak-magnitude}.
    
     To compute the completeness of the catalog from the source count distribution in Fig \ref{fig:Magnitude distribution plot FUV} and Fig \ref{fig:nuv source count}, we find the turnover point where the number count of the sources falls drastically. The results are given in Table \ref{table:magnitude-limit}. We see that the completeness increases in Bin3, i.e. with higher exposure time.

    In the NUV band, the N242W band with the highest number of detections has the turnover point around the magnitude $\sim22.602$.

\begin{figure*}
        \centering
        \begin{multicols}{2}
            \subcaptionbox{\label{fig:bin1 f1}}{\includegraphics[width=\linewidth]{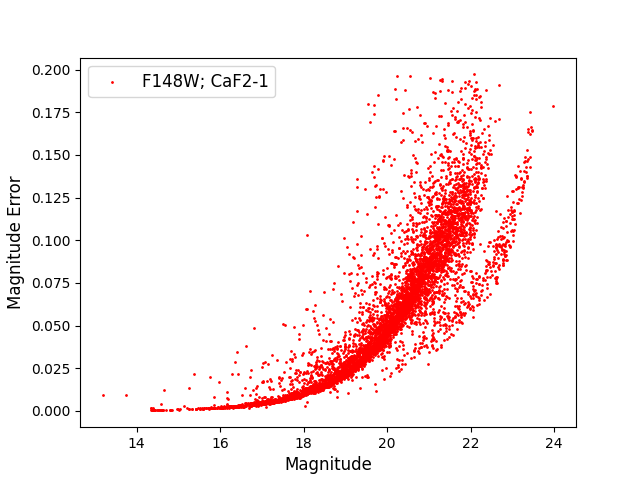}}\par 
            \subcaptionbox{\label{fig:bin1 f2}}{\includegraphics[width=\linewidth]{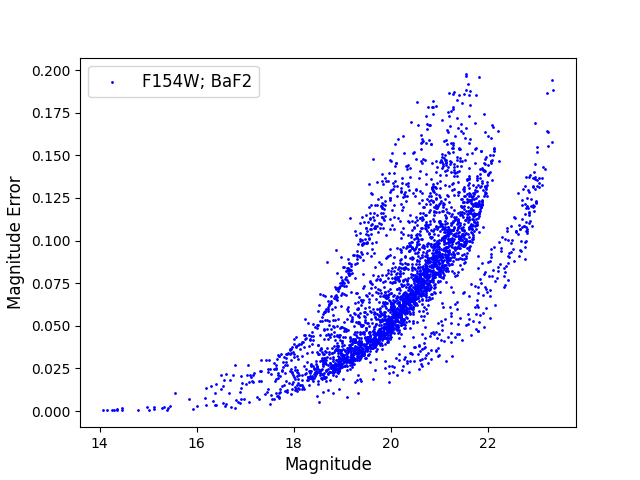}}\par 
        \end{multicols}
        \begin{multicols}{2}
            \subcaptionbox{\label{fig:bin2 f1}}{\includegraphics[width=\linewidth]{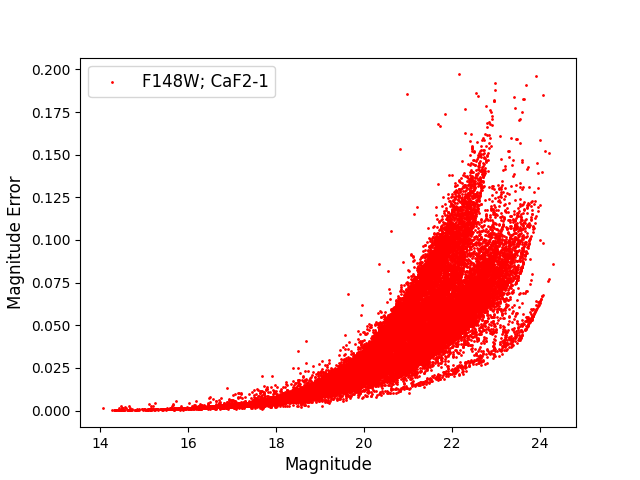}}\par
            \subcaptionbox{\label{fig:bin2 f2}}{\includegraphics[width=\linewidth]{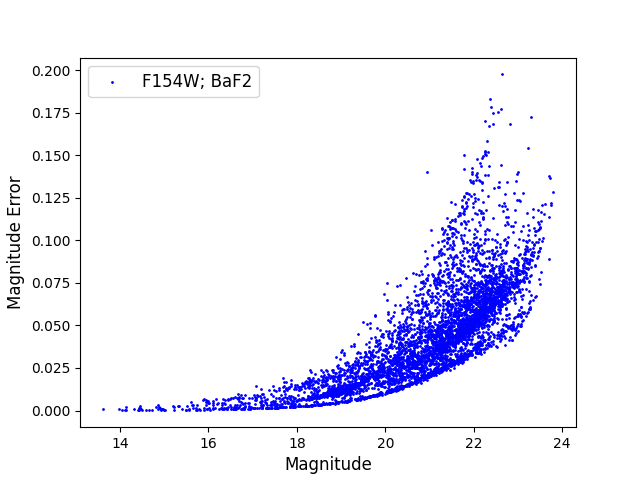}}\par
        \end{multicols}
        \begin{multicols}{2}
            \subcaptionbox{\label{fig:bin3 f1}}{\includegraphics[width=\linewidth]{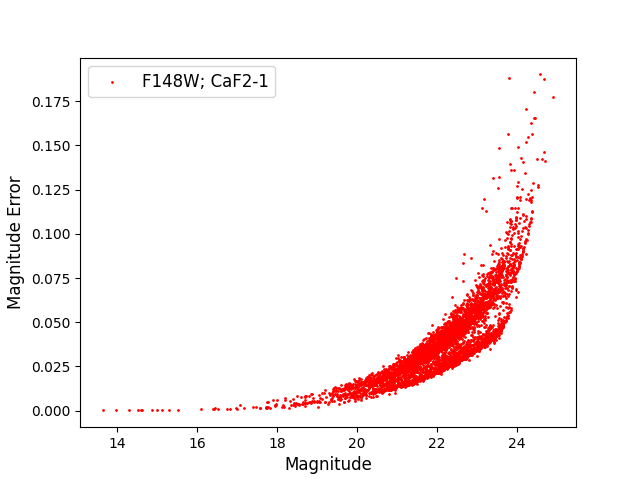}}\par
            \subcaptionbox{\label{fig:bin3 f2}}{\includegraphics[width=\linewidth]{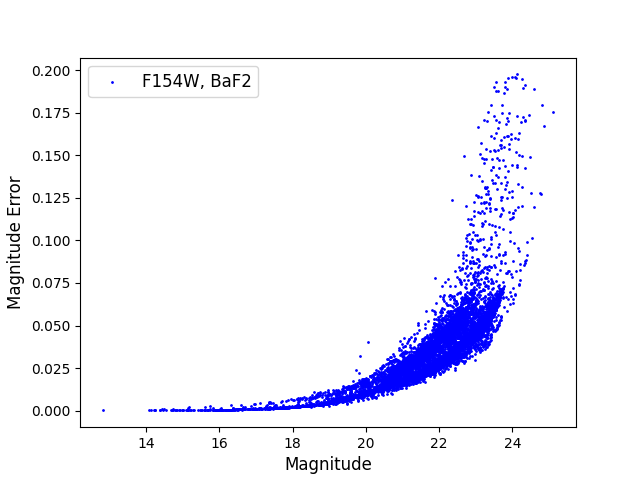}}\par
        \end{multicols}
    \caption{AB Magnitude vs AB Magnitude error for filters: FUV CaF2-1 (148W) and BaF2 (F154W) of UVIT. (a),(b): Bin 1, (c),(d): Bin 2 and (e),(f): Bin 3}
    \label{fig:fuvabmagerr}
    \end{figure*}
\begin{figure*}
        \centering
        \begin{multicols}{2}
            \subcaptionbox{\label{fig:magerr-c}}{\includegraphics[width=\linewidth]{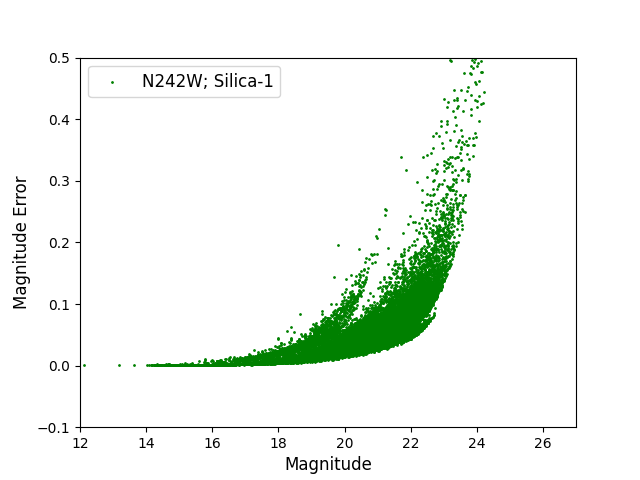}}\par
            \subcaptionbox{\label{fig:magerr-d}}{\includegraphics[width=\linewidth]{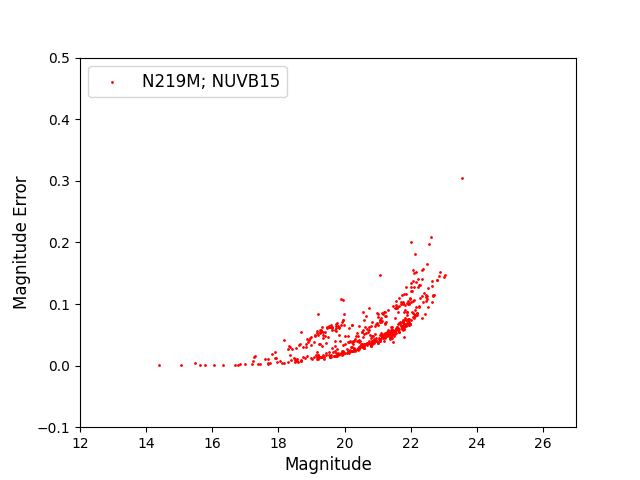}}\par
        \end{multicols}
    \caption{AB Magnitude vs AB Magnitude error for NUV filters: (a) NUV Silica-1 (N242W) and (b) NUVB15 (N219M) of UVIT}
    \label{fig:nuvabmagerr}
    \end{figure*}

    To determine the completeness from the magnitude vs magnitude error plot, we find the 5-$\sigma$ detection limit of the catalog's data. This corresponds to an error cut of 0.198; i.e. the faintest source which has a magnitude error of 0.198 or lower.
    The variation of magnitude error as a function of brightness (in AB magnitude) in the three bins is shown in Fig \ref{fig:fuvabmagerr}. For the F148W filter, the plot generated from each of the three bins shows the completeness at $\sim23.98$, 24.30, and 24.91 respectively as shown in Fig \ref{fig:bin1 f1}, \ref{fig:bin2 f1} and \ref{fig:bin3 f1}. For the second filter, we find the 5-$\sigma$ point at $\sim23.35$, 23.80, and 25.12 in all the three bins, as given in Fig \ref{fig:bin1 f2}, \ref{fig:bin2 f2} and \ref{fig:bin3 f2}. We find the limiting magnitude for the NUV N242W filter at about $\sim23.027$ (Fig \ref{fig:magerr-c}).

    As we have combined all the observations into a single catalog, we found that around $\sim1800$s exposure time, the highest number of observations has been made by UVIT. Hence, we can consider the magnitude limit of \textbf{Bin 1} as the completeness of the entire \textit{'UVIT-cat'} catalog, i.e. for FUV, we get the magnitude limit to be around $\sim21$, while for NUV, we find the completeness at around $\sim23$.

\begin{table}
        
        \resizebox{\linewidth}{!}{
        \begin{tabular}{cccccc}
            \hline
            Bins/Filter-slot & Filter-F1 & Filter-F2 & Filter-F3   & Filter-F5 & Filter-F7 \\

            \hline
            Bin1 & 20.527 & 20.568   & 20.065 & 19.523 & 21.474\\
            Bin2 & 20.818 & 21.967     & 21.499 & 21.093 & 20.49\\
            Bin3 & 21.98 & 22.907 & 22.032  & 20.614 & No data\\

        \end{tabular}}
        \caption{Peak AB magnitude of source-count distribution of each filter of FUV}
        \label{table:peak-magnitude}
    \end{table}

\begin{table}
        
        \resizebox{\linewidth}{!}{
        \begin{tabular}{cccccc}
            \hline
            Bins/Filter-slot & Filter-F1 & Filter-F2 & Filter-F3   & Filter-F5 & Filter-F7 \\

            \hline
            Bin1 & 21.283 & 21.033   & 20.587 & 19.97 & 21.561\\
            Bin2 & 21.945 & 22.681     & 22.499 & 21.939 & 22.295\\
            Bin3 & 23.671 & 23.769 & 22.822  & 22.356 & No data\\

        \end{tabular}}
        \caption{Magnitude limit of all the filters of FUV in three bins}
        \label{table:magnitude-limit}
    \end{table}

\begin{figure*}
        \centering
        \begin{multicols}{2}
            \subcaptionbox{GAIA EDR3\label{fig:ang gaia}}{\includegraphics[width=\linewidth]{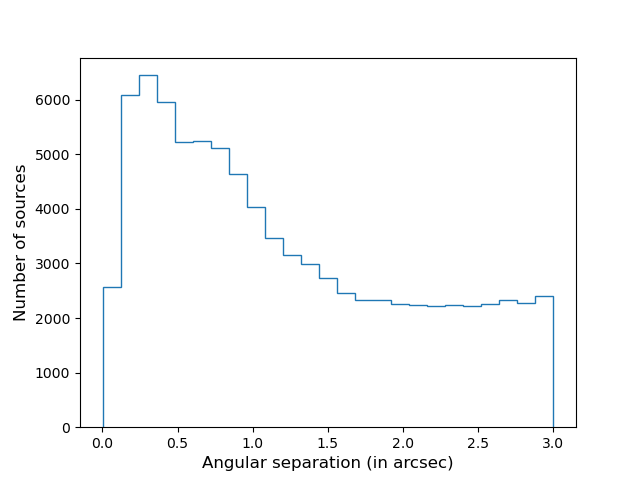}}\par 
            \subcaptionbox{GALEX\label{fig:ang galex}}{\includegraphics[width=\linewidth]{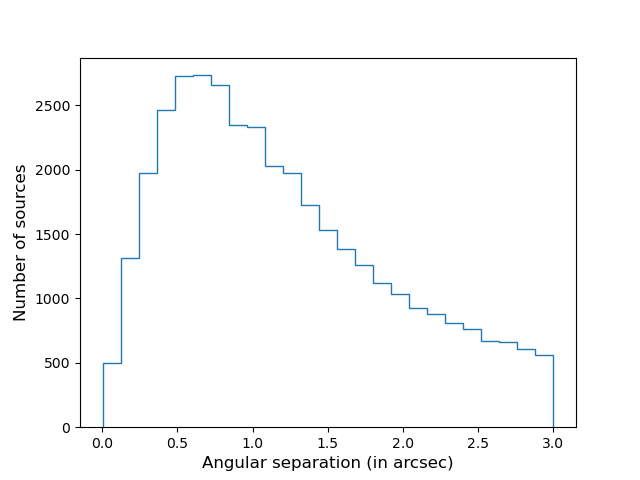}}\par
            
        \end{multicols}
    \caption{Angular separation distribution of UVIT with (a) GAIA EDR3 (b) GALEX}
    \label{fig:Ang sep}
    \end{figure*}

    We cross-matched the positions of our own UVIT catalog using the GALEX GUVcat and Gaia EDR3 catalog. For the following analysis, we plot and visualize the histogram distribution of angular separation of matched sources between the UVIT-cat catalog ($\alpha$,$\delta$) and the reference catalog position ($\alpha$,$\delta$) values respectively. We keep the separation threshold as 5\arcsec and find all matches, i.e. we consider the source as a match if it falls within this radius. This is shown in Fig \ref{fig:Ang sep}.

    We found the angular separation cross-matched with Gaia EDR3 to be $\simeq0.240\arcsec$, as given in Fig \ref{fig:ang gaia}. Similarly, we found the average angular separation when compared with GALEX's GUVcat to be $\simeq0.602\arcsec$, as given in Fig \ref{fig:ang galex}. This shows that our positions of the detected sources are in good agreement astrometrically with the already existing catalogs.

\section{Conclusion and Future Work}
    In our work, we have analysed 428 FUV and 54 NUV UVIT field pointings and present \textit{UVIT-cat} catalog. The UVIT-cat covers a total of $\sim63$ square degrees. We have come up with a catalog of $\simeq$102,773 sources in the 5 far ultraviolet and the near ultraviolet bands. The number of entries in each filter is tabulated in Table \ref{table:total_sources} and \ref{table:total_sources_nuv}. The magnitude limit after statistical analysis is found to be, $\sim21.28$, 21.03, 20.59, 19.523, and 21.474 for all FUV band filters respectively. The catalog contains 85 columns of data, and the description of the column keys is given in Table \ref{table:summary_table}. The catalog will help the scientific community study objects such as QSOs, white dwarfs, young galaxies, and various astrophysical sources and provide a clear picture of the sky in the ultraviolet band.  

\subsection{Acknowledgement}
    This publication uses the data from \textit{Indian Space Research Orgranisation's} (ISRO) \textit{AstroSat} mission. The data was available at the organization's archive \textit{Indian Space Science Data Centre}(ISSDC), where the data was processed at the Payload Operation Center (POC) at the \textit{Indian Institute of Astrophysics} (IIA). The UVIT mission is accomplished in collaboration between IIA, \textit{Inter-University Center for Astronomy and Astrophysics} (IUCAA), \textit{Tata Institute of Fundamental Research} (TIFR), ISRO and \textit{Canadian Space Agency} (CSA). The authors would like to thank the anonymous reviewers for their insightful reviews and comments. 

    \textit{Softwares:} IRAF \cite{irafpaper}, Astropy\cite{astropy}, Numpy\cite{numpy}, Pandas\cite{pandas}, Matplotlib\cite{matplotlib}, DS9, JUDE\cite{jude2017,jude2}, Astrometry.net \cite{Dustinastrometry}.

\textit{Data Availability:}  The complete electronic version of the UVIT-cat catalog in FITS format and the header files containing the WCS information can be found in the GitHub repository \footnote{\url{https://github.com/swagatastro98/UVIT-cat.git}}.

\bibliographystyle{unsrt}  
\bibliography{references}

\begin{thebibliography}{10}

\bibitem{Agrawal06}
P.~C. {Agrawal}.
\newblock {A broad spectral band Indian Astronomy satellite {\textquoteleft}Astrosat{\textquoteright}}.
\newblock {\em Advances in Space Research}, 38(12):2989--2994, January 2006.

\bibitem{kumar2012}
Amit Kumar, SK~Ghosh, J~Hutchings, PU~Kamath, S~Kathiravan, PK~Mahesh, J~Murthy, S~Nagbhushana, AK~Pati, MN~Rao, et~al.
\newblock Ultraviolet imaging telescope (uvit) on astrosat.
\newblock In {\em Space Telescopes and Instrumentation 2012: Ultraviolet to Gamma Ray}, volume 8443, pages 455--466. SPIE, 2012.

\bibitem{td1mission}
A.~Boksenberg, R.~G. Evans, R.~G. Fowler, I.~S.~K. Gardner, L.~Houziaux, C.~M. Humphries, C.~Jamar, D.~Macau, D.~Malaise, A.~Monfils, K.~Nandy, G.~I. Thompson, R.~Wilson, and H.~Wroe.
\newblock {The Ultra-violet Sky-Survey Telescope in the TD-1A Satellite}.
\newblock {\em Monthly Notices of the Royal Astronomical Society}, 163(3):291--322, 08 1973.

\bibitem{galexmission2005}
D~Christopher Martin, James Fanson, David Schiminovich, Patrick Morrissey, Peter~G Friedman, Tom~A Barlow, Tim Conrow, Robert Grange, Patrick~N Jelinsky, Bruno Milliard, et~al.
\newblock The galaxy evolution explorer: a space ultraviolet survey mission.
\newblock {\em The Astrophysical Journal}, 619(1):L1, 2005.

\bibitem{bianchi2017}
Luciana Bianchi, Bernie Shiao, and David Thilker.
\newblock Revised catalog of galex ultraviolet sources. i. the all-sky survey: Guvcat\_ais.
\newblock {\em The Astrophysical Journal Supplement Series}, 230(2):24, 2017.

\bibitem{leahy2020}
DA~Leahy, J~Postma, Y~Chen, and M~Buick.
\newblock Astrosat uvit survey of m31: point-source catalog.
\newblock {\em The Astrophysical Journal Supplement Series}, 247(2):47, 2020.

\bibitem{devaraj2023}
Ashish Devaraj, Prajwel Joseph, CS~Stalin, Shyam~N Tandon, and Swarna~K Ghosh.
\newblock Uvit observations of the small magellanic cloud: Point-source catalog.
\newblock {\em The Astrophysical Journal}, 946(2):65, 2023.

\bibitem{astrosatmission}
Kulinder~Pal Singh, SN~Tandon, PC~Agrawal, HM~Antia, RK~Manchanda, JS~Yadav, S~Seetha, MC~Ramadevi, AR~Rao, D~Bhattacharya, et~al.
\newblock Astrosat mission.
\newblock In {\em Space Telescopes and Instrumentation 2014: Ultraviolet to Gamma Ray}, volume 9144, pages 517--531. SPIE, 2014.

\bibitem{jude2}
P.~T. {Rahna}, Jayant {Murthy}, and Margarita {Safonova}.
\newblock {JUDE (Jayant's UVIT Data Explorer) pipeline user manual}.
\newblock {\em Journal of Astrophysics and Astronomy}, 42(2):35, October 2021.

\bibitem{Dustinastrometry}
Dustin Lang, David~W. Hogg, Keir Mierle, Michael~R. Blanton, and Sam~T. Roweis.
\newblock Astrometry.net: Blind astrometric calibration of arbitrary astronomical images.
\newblock {\em The Astronomical Journal}, 139:1782 -- 1800, 2009.

\bibitem{bertinsextractor}
Emmanuel Bertin and Stephane Arnouts.
\newblock Sextractor: Software for source extraction.
\newblock {\em Astronomy and astrophysics supplement series}, 117(2):393--404, 1996.

\bibitem{tandon2020additional}
SN~Tandon, J~Postma, P~Joseph, A~Devaraj, A~Subramaniam, IV~Barve, K~George, SK~Ghosh, V~Girish, JB~Hutchings, et~al.
\newblock Additional calibration of the ultraviolet imaging telescope on board astrosat.
\newblock {\em The Astronomical Journal}, 159(4):158, 2020.

\bibitem{jude2017}
J.~{Murthy}, P.~T. {Rahna}, F.~{Sutaria}, M.~{Safonova}, S.~B. {Gudennavar}, and S.~G. {Bubbly}.
\newblock {JUDE: An Ultraviolet Imaging Telescope pipeline}.
\newblock {\em Astronomy and Computing}, 20:120--127, July 2017.

\bibitem{irafpaper}
Doug Tody.
\newblock The iraf data reduction and analysis system.
\newblock In {\em Instrumentation in astronomy VI}, volume 627, pages 733--748. SPIE, 1986.

\bibitem{astropy}
Adrian~M Price-Whelan, Pey~Lian Lim, Nicholas Earl, Nathaniel Starkman, Larry Bradley, David~L Shupe, Aarya~A Patil, Lia Corrales, CE~Brasseur, Maximilian N{\"o}the, et~al.
\newblock The astropy project: sustaining and growing a community-oriented open-source project and the latest major release (v5. 0) of the core package.
\newblock {\em The Astrophysical Journal}, 935(2):167, 2022.

\bibitem{numpy}
Charles~R Harris, K~Jarrod Millman, St{\'e}fan~J Van Der~Walt, Ralf Gommers, Pauli Virtanen, David Cournapeau, Eric Wieser, Julian Taylor, Sebastian Berg, Nathaniel~J Smith, et~al.
\newblock Array programming with numpy.
\newblock {\em Nature}, 585(7825):357--362, 2020.

\bibitem{pandas}
Wes McKinney et~al.
\newblock pandas: a foundational python library for data analysis and statistics.
\newblock {\em Python for high performance and scientific computing}, 14(9):1--9, 2011.

\bibitem{matplotlib}
John~D Hunter.
\newblock Matplotlib: A 2d graphics environment.
\newblock {\em Computing in science \& engineering}, 9(03):90--95, 2007.

\end{thebibliography}

\end{document}